\documentclass{aa}  

\usepackage{txfonts}
\usepackage{graphicx}	
\usepackage{amsmath}	
\usepackage{amssymb}
\usepackage{caption}
\usepackage{placeins}
\usepackage{footnote}
\usepackage{float}
\usepackage[]{hyperref}%
\hypersetup{
    colorlinks,
    citecolor=blue,
    filecolor=magenta,      
    linkcolor=blue,
    }

\newcommand{\twelve}{$^{12}$}
\newcommand{\thirteen}{$^{13}$}

\newcommand{\plus}{$^{+}$}

\begin{document}

   \title{First detection of a deuterated molecule in a starburst environment within NGC 253}

   \author{
J. Butterworth\inst{\ref{inst.Leiden}} 
\and S. Mart\'in \inst{\ref{inst.ESOChile},\ref{inst.JAO}}
\and V. M. Rivilla \inst{\ref{inst.CAB-INTA}}
\and S. Viti \inst{\ref{inst.Leiden},\ref{inst.TRA},\ref{inst.UCL}} \and R. Aladro \inst{\ref{inst.Elsevier}}
\and L. Colzi \inst{\ref{inst.CAB-INTA}}
\and F. Fontani \inst{\ref{inst.INAF}, \ref{inst.MPE}, \ref{inst.LERMA}}
\and N. Harada \inst{\ref{inst.NAOJ},\ref{inst.ASIAA}, \ref{inst.SOKENDAI}}
\and C. Henkel \inst{\ref{inst.MPIfR},\ref{inst.Xinjiang}}
\and I. Jim\'enez-Serra \inst{\ref{inst.CAB-INTA}}
          }
\institute{\label{inst.Leiden}Leiden Observatory, Leiden University, PO Box 9513, NL-2300 RA Leiden, the Netherlands \\ \email{butterworth@strw.leidenuniv.nl }
\and\label{inst.ESOChile}European Southern Observatory, Alonso de C\'ordova, 3107, Vitacura, Santiago 763-0355, Chile  
\and\label{inst.JAO}Joint ALMA Observatory, Alonso de C\'ordova, 3107, Vitacura, Santiago 763-0355, Chile
\and\label{inst.CAB-INTA}Centro de Astrobiología (CAB, CSIC-INTA), Ctra. de Torrej\'on a Ajalvir km 4, 28850, Torrej\'on de Ardoz, Madrid, Spain
\and\label{inst.TRA}Transdisciplinary Research Area (TRA) `Matter'/Argelander-Institut für Astronomie, University of Bonn
\and\label{inst.UCL}Physics and Astronomy, University College London, UK
\and\label{inst.Elsevier}Elsevier, Radarweg 29, 1043 NX Amsterdam, Netherlands
\and\label{inst.INAF}Istituto Nazionale di Astrofisica (INAF), Osservatorio Astrofisico di Arcetri, Florence, Italy
\and\label{inst.MPE}Max-Planck-Institute for Extraterrestrial Physics (MPE), Garching bei M\"unchen,
Germany
\and\label{inst.LERMA}Laboratoire d’\'Etudes du Rayonnement et de la Mati\`ere en Astrophysique et
Atmosph\`eres (LERMA), Observatoire de Paris, Meudon, France
\and\label{inst.NAOJ}National Astronomical Observatory of Japan, 2-21-1 Osawa, Mitaka, Tokyo 181-8588, Japan
\and\label{inst.ASIAA}Institute of Astronomy and Astrophysics, Academia Sinica, 11F of AS/NTU Astronomy-Mathematics Building, No.1, Sec. 4, Roosevelt Rd, Taipei 10617, Taiwan
\and\label{inst.SOKENDAI}Department of Astronomy, School of Science, The Graduate University for Advanced Studies (SOKENDAI), 2-21-1 Osawa, Mitaka, Tokyo, 181-1855 Japan
\and\label{inst.MPIfR}Max-Planck-Institut f\"ur Radioastronomie, Auf-dem-H\"ugel 69, 53121 Bonn, Germany    
\and\label{inst.Xinjiang}Xinjiang Astronomical Observatory, Chinese Academy of Sciences, 830011 Urumqi, China
}

   \date{Received ; accepted }

 
  \abstract
   {Deuterium was primarily created during the Big Bang Nucleosynthesis (BBN). This fact, alongside its fractionation reactions resulting in enhanced abundances of deuterated molecules, means that these abundances can be used to better understand many processes within the interstellar medium (ISM), as well as its history. Previously, observations of deuterated molecules have been limited to the Galaxy, the Magellanic Clouds and (with respect to HD) to quasar absorption spectra.}
   {We present  the first robust detection of a deuterated molecule in a starburst environment and, besides HD, the first one detected outside the Local Group. We therefore can constrain the deuterium fractionation, as observed by DCN.}
   {We observed the CMZ of the nearby starburst galaxy NGC 253 covering multiple Giant Molecular Clouds (GMC) with cloud scale observations ($\sim 30$ pc) using ALMA. Via the use of the \texttt{MADCUBA} package we were able to perform LTE analysis in order to obtain deuterium fractionation estimates.} 
   {We detect DCN in the nuclear region of the starburst galaxy NGC 253 and estimate the deuterium fractionation (D/H ratio) of DCN within the GMCs of the CMZ of NGC 253. We find a range of $5 \times 10^{-4}$ to $10 \times 10^{-4}$, relatively similar to the values observed in warm Galactic star-forming regions. We also determine an upper limit of D/H of $8 \times 10^{-5}$ from DCO\plus within one region, closer to the cosmic value of D/H. }
   {Our observations of deuterated molecules within NGC 253 appear to be consistent with previous galactic studies of star forming regions.  
   This implies that warmer gas temperatures increase the abundance of DCN relative to other deuterated species. This study also further expands the regions, particularly in the extragalactic domain, in which deuterated species have been detected.}

   \keywords{Interstellar medium (ISM): molecules, galaxies: active - starburst - ISM, astrochemistry.
               }

   \maketitle
%

\section{Introduction}
\label{sec:Int}

Deuterium, one of the two stable isotopes of hydrogen, was primarily synthesized during the Big Bang Nucleosynthesis, which occurred during the first seconds after the Big Bang \citep{1976Epstein,2003Prodanovic,2016Cyburt}.
Deuterium has predominantly been observed in the form of molecules in dense Galactic star-forming cloud cores, and the abundance of these deuterated molecules with respect to their hydrogenated counterparts \citep[D/H  $\sim10^{-3}-10^{-1}$][]{Crapsi2005,2008caselli,2022_Colzi} exceeds by orders of magnitude the [D/H] elemental abundance in the Solar neighbourhood \citep[$\sim 10^{-5}$][]{2003Oliveira}. Since stars are not a significant source of deuterium, and because deuterium is destroyed in their interior, its abundance, i.e. the D/H ratio, is expected to decrease with cosmic time \citep{Fontani2011}. In fact, observed D-abundance within molecules has been shown to be enhanced due to ongoing chemical processes 
\citep{2012Caselli}. 

Deuterium fractionation can  result from the formation processes of different deuterated species, which  depend on the physical conditions of a region \citep{2018_Colzi,2022_Colzi}. This fractionation can then result in a column density ratio between a species containing D and its hydrogenated counterpart ($D_{frac}$) higher than the primodial value. As an example, DCN has multiple formation pathways with the primary pathway (62\%) being $\text{HD} + \text{CH}_3^+ \rightarrow \text{CH}_2\text{D}^+$ which then forms CHD and $\text{CH}_2\text{D}$, subsequently reacting with N to form DCN \citep{2001_Turner}. A secondary formation mechanism (22\%) of DCN includes the reaction, $\text{HD} + \text{H}_3^+ \rightarrow \text{H}_2\text{D}^+$. However, $\text{H}_2\text{D}^+$ is also consumed in the sole formation pathway of the deuterated cations N$_2$D\plus and DCO\plus. These cations are formed most abundantly at low temperatures (T $< 20$K) where the ratio of H$_2$D+/H$_3^+$ is enhanced \citep{1987_Wootten}. Notably, deuterium fractionation ($D_{frac}$) in galactic massive dense cores exhibits values ranging from $\sim0.01$ to $0.7$, substantially exceeding the primordial D/H ratio  \citep[][]{Fontani2011,Fontani2015,2017_Zahorecz}. The [D/H] molecular ratio has been shown to decrease with increasing dust temperature in N$_2$D\plus and its hydrogenated counterpart by \cite{Emprechtinger2009}. Thus, the deuteration fraction becomes a powerful tool to estimate the time since massive stars started heating their medium \citep{Fontani2014}. This fraction may probe the degree of processing of the gas by star formation.
DCN on the other hand is primarily enhanced in warmer gas \citep[up to $\sim 70~K$.][]{2013_Roueff,Fontani2015,Gerner2015}. Thus, processes that influence the temperature of an environment, such as shocks may also affect the deuterium fractionation. This is seen in the the deuterium fractionation of DCN/HCN within shocked galactic regions like L1157-B1, which are associated with particularly high DCN/HCN abundance ratios \citep{2017_Busquet}.

Lyman Limit System (LLS) measurements, in the line of sight high-z quasars ($z\sim2.5$) indicate a primordial D/H ratio of around $2.55 \pm (0.03) \times 10^{-5}$ \citep{2018Zavarygin}. In the Galactic Center (GC), where matter has undergone particularly intense stellar processing, even lower D/H ratios are expected. The GC also typically possesses high kinetic temperatures (T$_{kin}$), thus deuteration is not expected to be efficient.
There are 2 studies that have derived the D/H ratio within the Galactic center.
Firstly, a study by \citet{Lubowich2000} identified a deuterium to hydrogen of ($D/H \sim 1.7 \times 10^{-6}$) within the Galactic bulge; this value was five orders of magnitude above the value of $5\times10^{-12}$ within the central bulge of the Milky Way, as predicted by nucleosynthesis models within this region.
Such enhancement was claimed to result from pristine material feeding the central molecular zone of our Galaxy.
Surprisingly, however,  measurements in GC star-forming regions report [D/H] values of $2-4 \times 10^{-4}$, derived from deuterated molecular species \citep{2022_Colzi}. In this work the authors used the high D/H ratio to pinpoint positions in the GC where a future generation of stars will form.

Despite extensive studies in the Milky Way, extragalactic investigations of deuterium fractionation remain limited due to observational challenges. Notable exceptions include detections of deuterated molecular hydrogen (HD) in absorption towards damped Lyman-$\alpha$ systems \citep{Ivanchik2010}. There are also reported detections of deuterated species (such as DCO\plus) towards  star-forming regions in the Large Magellanic Cloud (LMC)  with an estimated D/H ratio of $\sim 10^{-4}$ \citep{Chin1996,Heikkilae1997}. However, the information towards even the nearest and brightest galaxies consists of upper limits or tentative detections of DCN towards a couple of sources \citep{Mauersberger1995}.
In particular,  DCN and N$_2$D\plus have been tentatively observed towards the starbursting environment of NGC 253 by \cite{Mart'in2006}.
\cite{Mart'in2006} obtained an upper limit for DCN, DCO$^+$ and a tentative detection of N$_2$D$^+$ resulting in a D/H ratio of $<1-4 \times 10^{-3}$ for these molecules at low spatial resolution ($\sim 300$~pc).

In this paper, we present the first robust detection of a deuterated molecule in a non-Magellanic Cloud extragalactic starburst environment, the starburst galaxy NGC 253, which is (besides HD) also the first detection of such a molecule outside the Local Group. We hence derive for the first time a robust D/H ratio and compare it with both galactic ratios and those observed within the LMC, as well as with theoretical models. By examining deuterated species, we seek to uncover clues about the evolutionary state of star-forming regions and the impact of galaxy-scale processes on deuterium enrichment.

\section{Observations}
\label{sec:Obs}
We carried out interferometric observations using the
Atacama Large Millimeter/Submillimeter Array (ALMA) in Cycle 5 in the period between June and August 2018 as part of the project 2017.1.00028.S. 
The observations were centered at the coordinates $\alpha_{J2000.0}$= 0$^h$ 47$^m$ 33.182$^s$ and $\delta_{J2000.0}$= -25$^{\circ}$ 17$^{\prime}$ 17.148$^{\prime\prime}$. 
We used two different spectral setups: one in Band 4 to target the $J$=2$-$1 transitions of the deuterated species DCN, DCO$^{+}$, and N$_2$D$^+$, and one in Band 5 to target the same transition of the hydrogenated species H$^{13}$CN, H$^{13}$CO$^{+}$, and N$_2$H$^+$ (see Table \ref{tab:transitions}). The information of the observations in each band (date, number of antennas, range of baselines, time on source, rms, system temperatures and precipitable water vapour) are summarized in Table \ref{tab:observations}.

The data was calibrated and imaged using standard ALMA calibration scripts of the Common Astronomy Software Applications package (CASA) \citep{2022_CASA}\footnote{https://casa.nrao.edu}. 
We used the \texttt{CASA} task {\it tclean}, with the {\it auto-multithresh} option, which automatically masks regions during the cleaning process. We chose a common beam for all the spectral datacubes on Band 4 and 5 of 1.6"$\times$1.2" and PA=80$^{\circ}$. 
The continuum was substracted in the image plane using the STATCONT\footnote{STATCONT is a python-based tool designed to determine the continuum emission level in line-rich spectral data; https://hera.ph1.uni-koeln.de/~sanchez/statcont} package (\citet{sanchez-monge2017}).
The expected flux density calibration uncertainty is of 5$\%$, as estimated from recent analysis of calibrators in bands 3 and 6 \citep[][and references therein]{bonato2018,2019ALMA}.
The integrated spectral line intensities from our data cubes were extracted using \texttt{CubeLineMoment}\footnote{\url{https://github.com/keflavich/cube-line-extractor}} \citep{2019_Mangum_Heart}. 
\texttt{CubeLineMoment} works by extracting integrated intensities for a given list of targeted spectral frequencies by applying a set of spectral and spatial masks (defined by the user).
  The \texttt{CubeLineMoment} masking process uses a brighter spectral line (typically the main isotopologue line of the same rotational transition), whose velocity structure over the galaxy is most representative of the science target line, which is used as a tracer of the velocity of the 
inspected gas component. The products of \texttt{CubeLineMoment} are moment 0, 1, and 2 maps in the  units chosen by the user, masked below a chosen $\sigma$ threshold (channel-based). 

%
%
%
%
%
%
%
%



\begin{figure}
  \begin{tabular}[b]{@{}p{0.5\textwidth}@{}}
    \centering\includegraphics[width=1.0\linewidth]{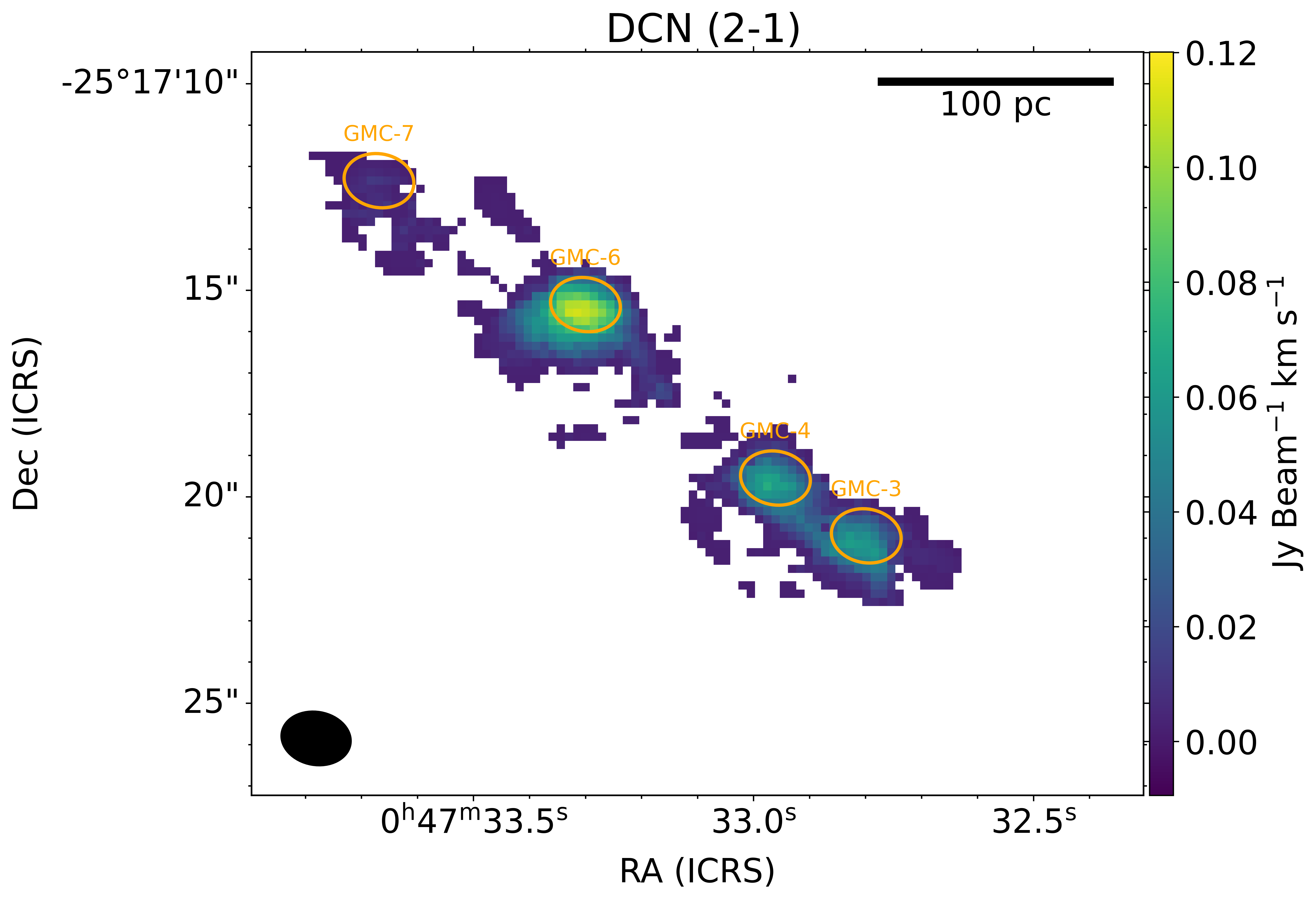} 

    \centering\small (a)
      \end{tabular}%
  \quad
  \begin{tabular}[b]{@{}p{0.5\textwidth}@{}}
    \centering\includegraphics[width=1.0\linewidth]{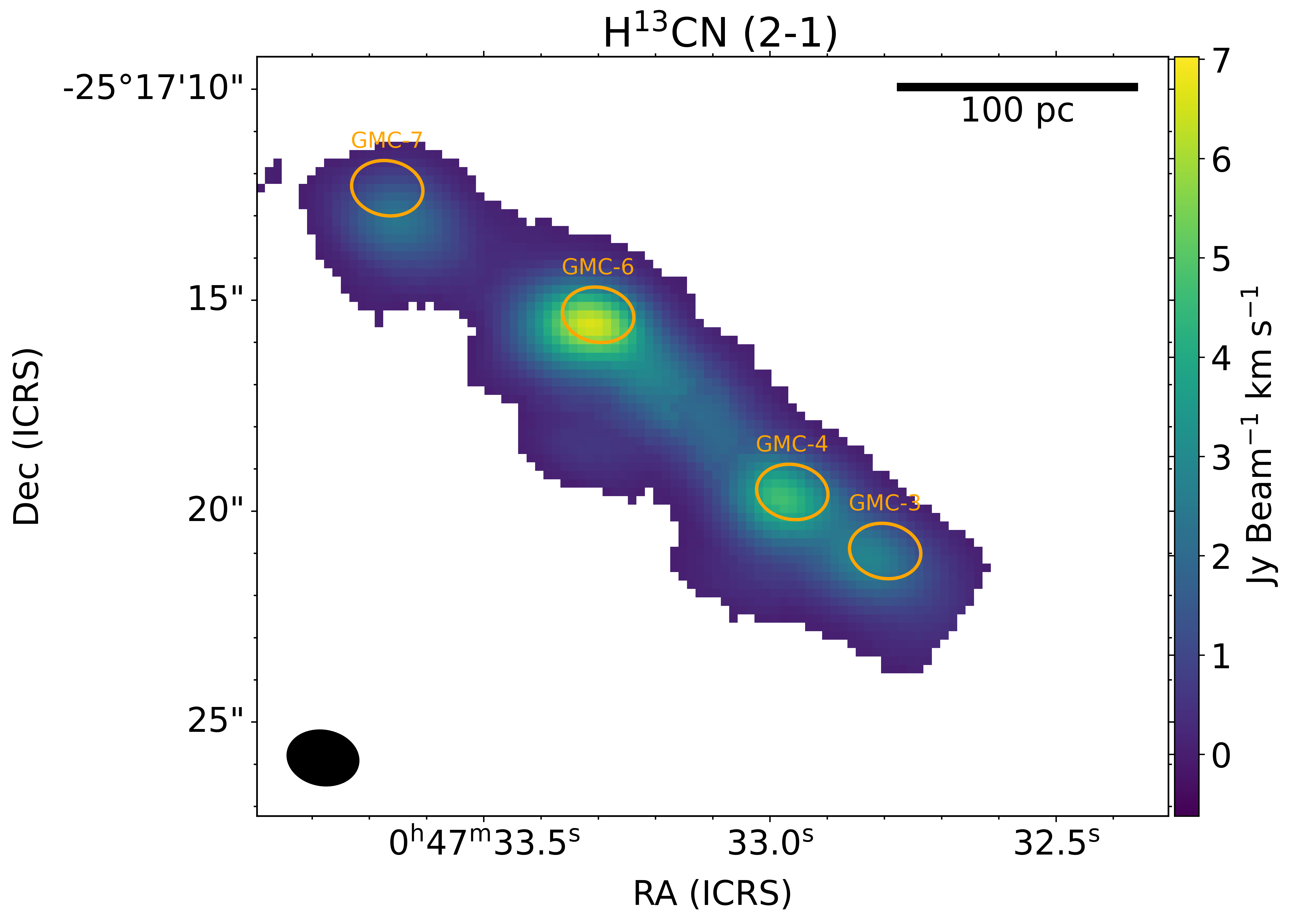} 

    \centering\small (b) 
  \end{tabular}%
\caption{Velocity-integrated line intensity moment 0 maps, given in [Jy\,km\,s$^{-1}$ /beam], for the observed DCN (2-1) line (a) alongside the accompanying H\thirteen CN (2-1) line (b). Both of these maps are shown with a 3 sigma cut off. The orange ellipses show the beam-sized regions over which the later analysis of this letter was conducted.
  }
  \label{fig:mom0_DCN}
\end{figure}

\begin{figure}
    \centering
    \includegraphics[width=1.0\linewidth]{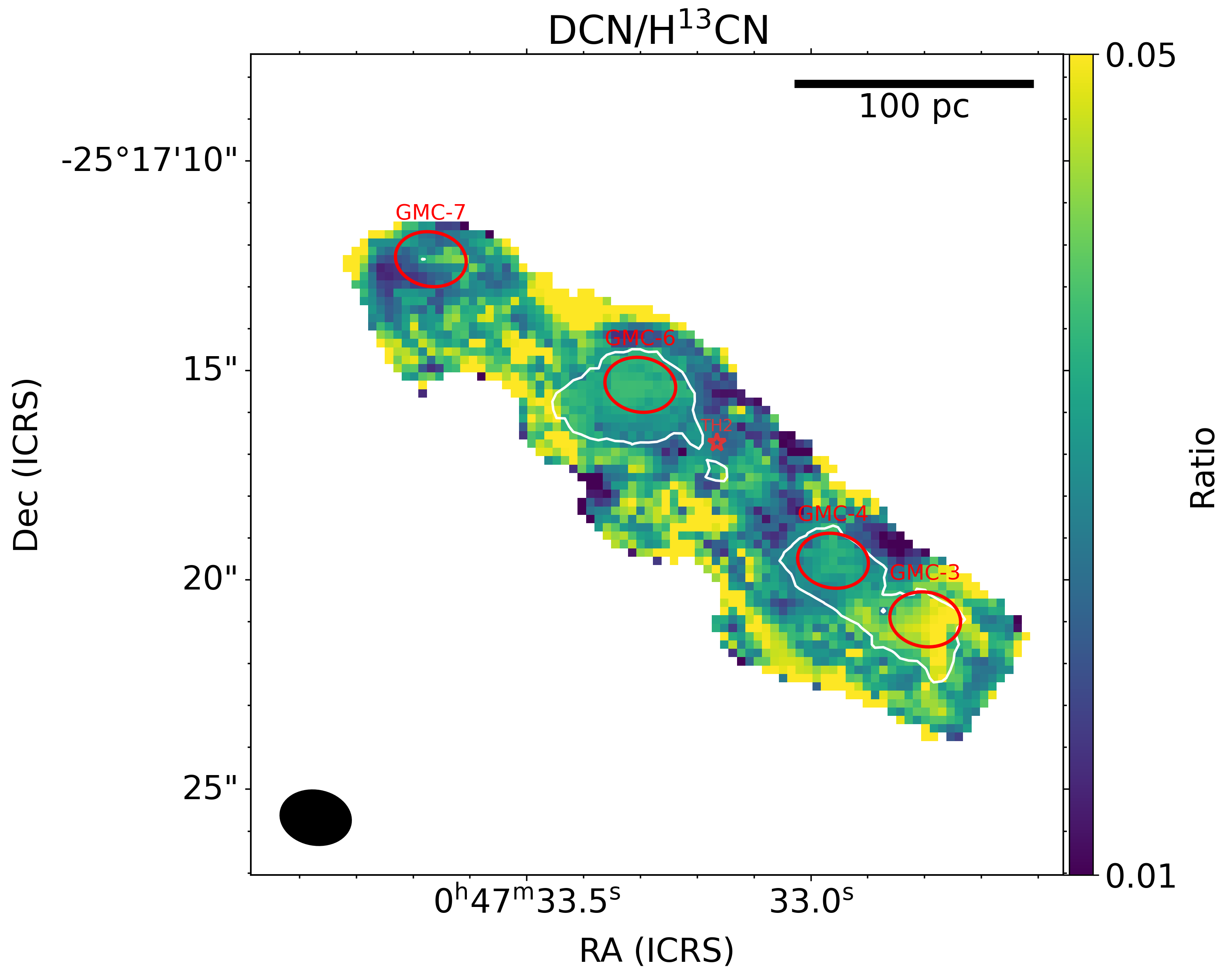}
    \caption{The ratio map of the velocity integrated line intensities of DCN (2-1)/H\thirteen CN (2-1). The white contours shown are the 5 sigma limited intensities of DCN (2-1). }
    \label{fig:12_13C_map}
\end{figure}

\begin{figure}
  \centering
  \includegraphics[width=0.5\textwidth]{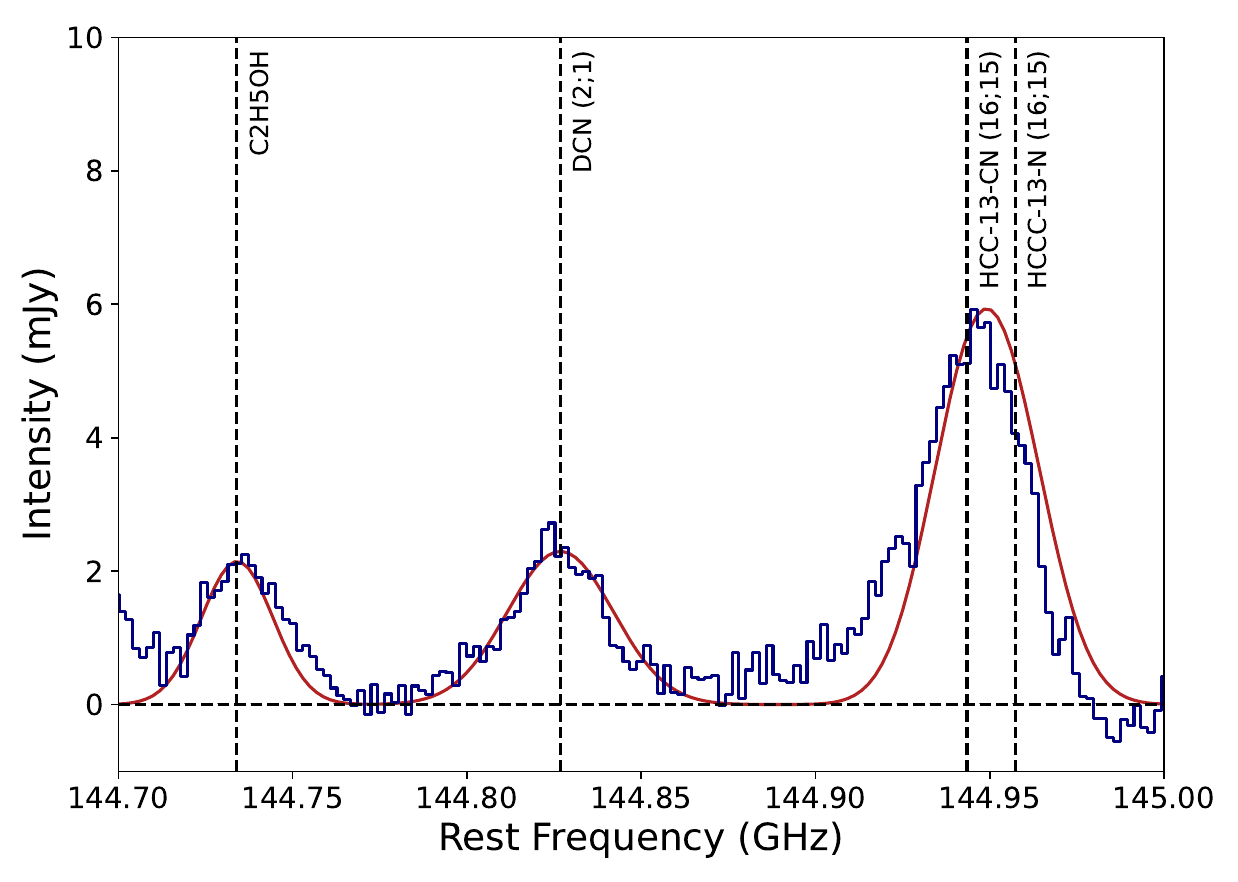}
  \caption{DCN (2-1) spectrum observed in the brightest region GMC-6 (in blue). The best fitting LTE model from \texttt{MADCUBA} is shown overplotted in red. }
  \label{fig:DCN_Bright_Spec}
\end{figure}

\begin{table}
\centering
\caption{Spectroscopy of the deuterated molecular transitions and of their hydrogenated counterparts studied in this work from CDMS \citep{CDMS_2001,CDMS_2005,CDMS_2016}. The spectroscopic information of additional lines used in this paper are given in Appendix \ref{app:Spec_Info}.}
\begin{tabular}{c c c c c}
\hline
Molecule & Frequency & Transition  & logA$_{\rm ul}$  & E$_{\rm up}$  \\
 & (GHz) &   &  (s$^{-1}$) & (K)  \\
\hline
DCN & 144.8280015 & 2$-$1 & -3.89786  & 10.4  \\
H$^{13}$CN & 172.6778512 & 2$-$1 & -3.67033  & 12.4   \\
\hline
DCO$^+$& 144.0772890 & 2$-$1 & -3.18099  &  10.4  \\
H$^{13}$CO$^+$& 173.5067003 & 2$-$1 & -3.43195  &  12.5  \\
\hline
N$_2$D$^+$& 154.2170318 & 2$-$1 &  -3.70470 & 11.1   \\
N$_2$H$^+$ & 186.3446844 &2$-$1  & -3.45807  & 13.4   \\
\hline 
\end{tabular}
\label{tab:transitions}
\end{table}

\begin{table*}
\centering
\caption{The parameters of the ALMA observations used as a part of this study.}
\begin{tabular}{c c c c c c c c}
\hline
Band & Date & $\#$   & Baselines  & $t_{\rm OS}^{(a)}$  & $rms^{(b)}$ & $T_{\rm sys}$ & $pwv^{(c)}$  \\
 & 2018 & Antennas  & (m)  &  & (mJy/beam) &  (K)  & (mm) \\
\hline
4 & June $\&$ Aug.  & 46$-$48  & 15$-$782  & 4.5 h & 0.38 & 52$-$90  & 0.4$-$4.6 \\
5 & Aug.  &  46 & 15$-$500 & 15 min & 0.56 & 140  & 0.5 \\ 
\hline
\end{tabular}\\
\footnotesize{(a) Time on the source NGC 253. (b) Measured in 10 km s$^{-1}$. (c) Precipitable water vapour.}
\label{tab:observations}
\end{table*}


\section{Analysis and results}
\label{sec:ResultsandAnalysis}
In this paper we present the first observed spectra of a deuterated species detected within an extra-galactic starburst environment. Using  observations of deuterated DCN in combination with the hydrogenated isotopologue H\thirteen CN we have computed an estimate of the D/H ratio towards four GMC regions within the NGC 253 central molecular zone. These GMC regions are based upon both continuum and molecular line emission observations previously defined within NGC 253 \cite[e.g.][]{2015_Leroy,2018_Leroy,2022_Levy}. Specifically, the four innermost GMCs are studied, GMC-4, 5, 6 and 7. The detections of DCN within these regions are provided alongside upper limits of the deuterium fractionation from the deuterated species DCO$^+$ and N$_2$D$^+$. The integrated intensities of each line within each region are shown in Table \ref{tab:Intensities}. In Figure~\ref{fig:mom0_DCN} we present the velocity-integrated line intensity moment 0 maps from DCN (2-1) and H \thirteen CN (2-1) obtained via \texttt{CubeLineMoment}. The moment 0 maps of both N$_2$H$^+$ (2-1) and H\thirteen CO\plus, alongside the moment 0 maps of the two HC$_3$N lines used as a part of this study are shown in Appendix \ref{app:Mom0maps}. The ratio map of DCN/H\thirteen CN is given in Figure \ref{fig:12_13C_map}. 

To conduct our analysis we have used the SLIM (Spectral Line Identification and Modeling) tool within the MADCUBA package \citep{2019MartinMADCUBA} \footnote{Madrid Data Cube Analysis on ImageJ which is a software developed in the Center of Astrobiology (CAB, CSIC-INTA) to visualize and analyze astronomical single spectra and datacubes. 
MADCUBA is available at \hyperlink{https://cab.inta-csic.es/madcuba/}{https://cab.inta-csic.es/madcuba/}} which allows us to identify and perform  multi-transition profile fitting. To accomplish this, SLIM generates a synthetic spectrum, assuming local thermodynamic equilibrium (LTE) conditions, and 
finds the best nonlinear least-squares fit to the data as well as the associated statistical errors. The free parameters to be fitted are the column density of the molecule, $N_{\text{mol}}$, the excitation temperature, $T_{\text{ex}}$, the peak velocity, v$_\text{LSR}$, and the full width at half maximum (FWHM).
In this paper, due to being limited to a single line in both DCN and H\thirteen CN, the parameter that was left free was the column density $N_{\text{mol}}$. In order to derive the temperature within each region we decided to use the serendipitous detection of multiple lines of CH$_3$CCH. CH$_3$CCH is a molecule with a known population dependence on kinetic temperature. As a result of it being a symmetric top molecule CH$_3$CCH possesses a dipole moment in line with its carbon chain. This means that transitions with $\Delta K > 0$ are forbidden (where $k$ is the projection of the total angular momentum, $J$, on the rotational axis). Thus, different $K$-components are connected only via collisional processes, making their relative population sensitive to the kinetic temperature \citep{2015_Fayolle,2019_Calcutt,2024_Khalifa}. Due to the complex hyperfine structure of CH$_3$CCH we were able to observe multiple lines of the $J$ = 9--8 and $J$ =11--10 transitions. The temperatures for each region are given in Table \ref{tab:Ratios}. The resulting $T_{ex}$  from CH$_3$CCH has been shown to be a good estimate of the kinetic temperature of the gas at densities of n$_H \geq 10^{3-4}$ cm$^{-3}$ \citep{Ch3cch_1,Ch3cch_2}. The source size of each region was assumed to be equal to the beam size, and thus the beam-sized region intensities were extracted.
To perform the LTE line fitting we use the \texttt{MADCUBA} AUTOFIT function which provides the best solution for the free parameters, and their associated errors. Also, H\thirteen CN has been found to be optically thin within NGC 253, and thus likely DCN too, at similar regions at the similar spatial scales \citep{2024_Butterworth}.This was tested in \texttt{MADCUBA}, varying the the T$_{ex}$ within the LTE modelling had a negligible effect

To determine the D/H ratios in each region the column density of H\thirteen CN has been converted to the column density of the primary isotopologue (HCN) by determining the \twelve C/\thirteen C ratio from the simultaneously observed HC$_3$N and two of its prominent isotopologues, HC\thirteen CCN and HCC\thirteen CN. The calculated \twelve C/\thirteen C ratios ranged from $\sim$40-56 across the 4 regions. These values are on the high end but are relatively consistent with the ratio derived by \cite{2021Martin} using HC$_3$N at 16" resolution towards NGC 253; they are also consistent  with the \twelve C/\thirteen C ratio obtained by \cite{2019_Tang} using ALMA observations of CN and its \thirteen C-bearing isotopologue within NGC 253. Table \ref{tab:Ratios} lists the obtained D/H ratios for each GMC region as indicated by DCN as well as the estimated upper limits for N$_2$D$^+$ and DCO$^+$. DCN (2-1) benefits from not having many nearby lines potentially contaminating the detection. Thanks to the broadband spectral scan carried out by ALMA (ALCHEMI, \citealt{2021Martin}) we can safely confirm the lack of contamination by known species. Of known lines at the same frequency as DCN (2-1) at the redshift of NGC~253, they are only of either complex species not previously observed in extragalactic sources, or low intensity hyperfine structure lines from species such as CH$_3$OH \citep{CDMS_2016}. For each of these cases of unknown lines we can say with some assurance that they are not contaminating the DCN (2-1) detection within our regions. While the low resolution data published by \cite{2021Martin} does not indicate the presence of DCN, the full spectral modelling of the high resolution data (Lopez-Gallifa in prep.) shows hints of the emission of DCN in four different transitions from 144 GHz to 362 ~GHz. The transition reported here is the only one not blended with emission from other species as shown in Fig. \ref{fig:DCN_Bright_Spec}.
We note that the data presented here are much deeper than those reached in the ALCHEMI data.  The spectra of all four GMC regions for all investigated lines are given in Appendix \ref{app:Spectra}.



\begin{figure*}
  \centering
  \includegraphics[width=1.0\textwidth]{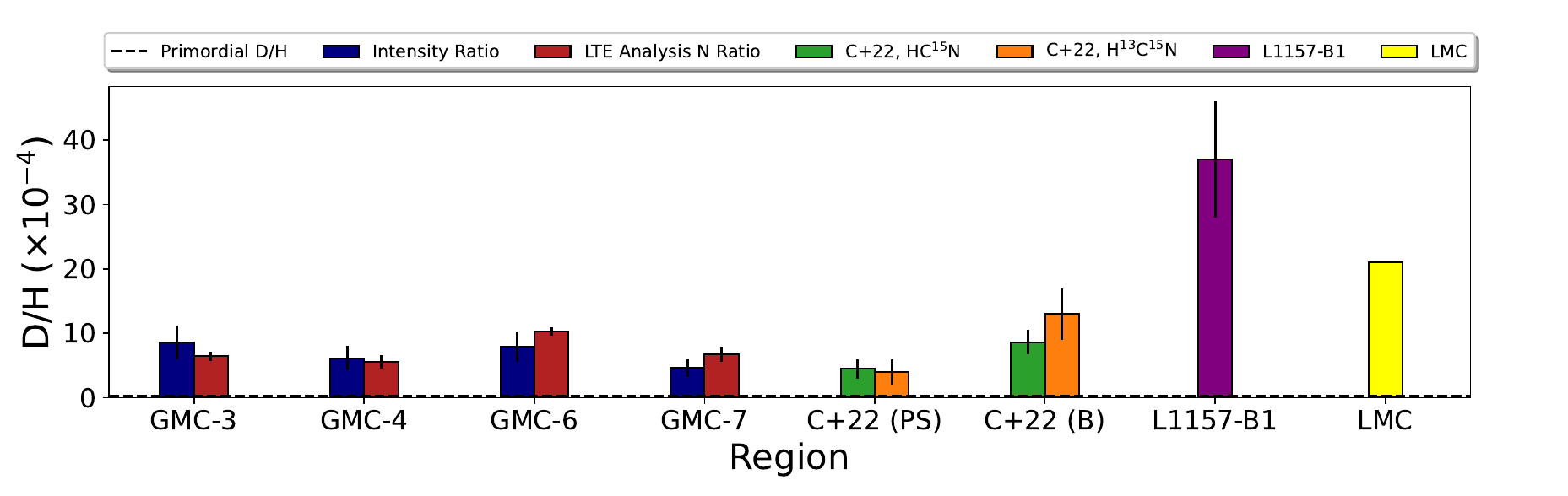}
  \caption{The D/H as obtained through DCN for each region across NGC 253, shown both for the intensity ratio of DCN (2-1)/H\thirteen CN (2-1) (in blue) and column density ratio (in red). These observations are shown relative to the prestellar (PS) and broad (B) components of G+0.693-0.027 from \cite{2022_Colzi} (C+22, shown in green and orange, respectively), the D/H ratio observed in the shock region L1157-B1 by \cite{2017_Busquet} (in purple) and the DCN/HCN ratio observed in the LMC by \cite{Chin1996} (in yellow).  }
  \label{fig:Swarm}
\end{figure*}

\begin{figure*}
  \centering
  \includegraphics[width=1.0\textwidth]{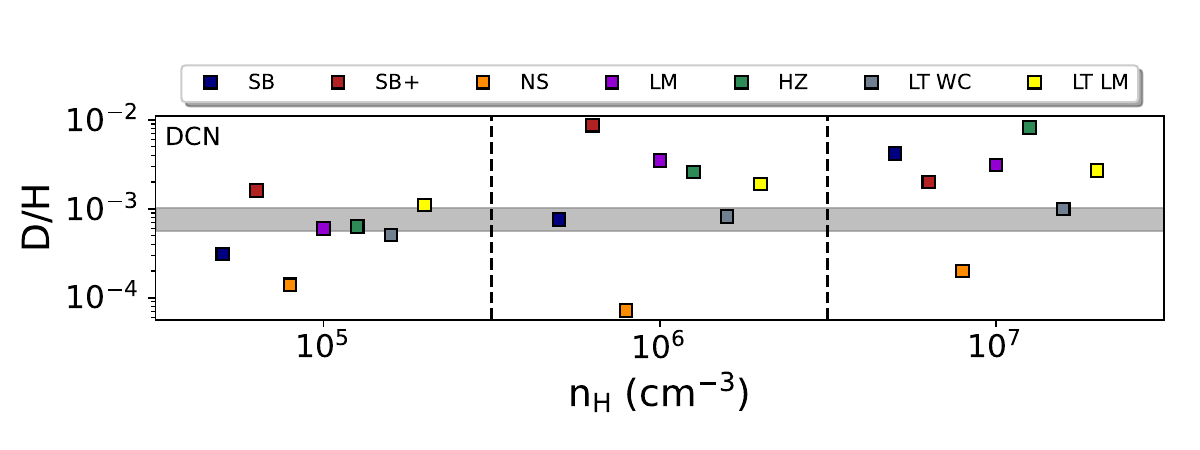}
  \caption{Comparison of LTE analysis Predicted Column density ranges of DCN/HCN to models shown in \cite{2007_Roueff} and \cite{Bayet2010}. SB stands for Starburst, SB+ for Cosmic-ray enhanced starburst, NS for Normal Spiral, LM for Low Metallicity, HZ for High redshift and WC for Warm Core. The LT (Lower Temperature) prefix is given to the models from \cite{2007_Roueff} as these models were conducted at a significantly lower kinetic temperature (70 K), than the remaining models from \cite{Bayet2010} (>300 K). The effect of this temperature difference is discussed in the text. The primordial D/H ratio of around $2.55 \pm (0.03) \times 10^{-5}$ \ is presented with a horizontal dashed line \citep{2018Zavarygin}.}
  \label{fig:Model_DCN}
\end{figure*}

\begin{figure}
  \centering
  \includegraphics[width=0.5\textwidth]{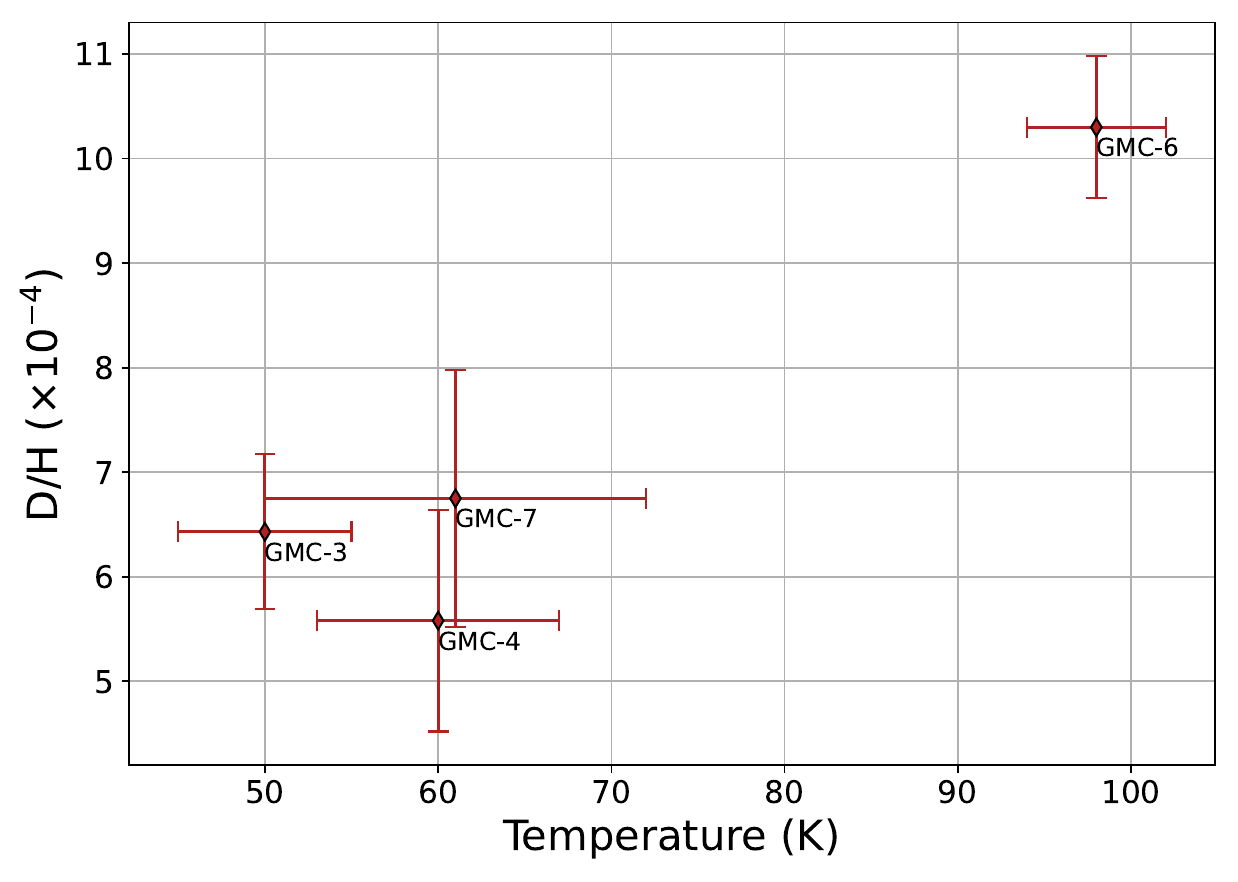}
  \caption{Derived excitation temperatures of DCN from LTE analysis versus the predicted deuterium fractionation derived for DCN across the GMC regions. }
  \label{fig:DCN_Temp}
\end{figure}


\begin{table*}[ht!]

\centering
\parbox{\textwidth}{\caption{Integrated intensities in [K km s$^{-1}$] as observed within each of the studied GMC regions. The errors presented represent 1$\sigma$. The integrated intensities are provided only for the non-contaminated lines.} \label{tab:Intensities}}

\begin{tabular}{cc|cccc}
               &  & \multicolumn{4}{c}{Intensities (K km s$^{-1}$)} \\ \hline
Molecule  & Transition       & GMC-3         & GMC-4          & GMC-6          & GMC-7         \\ \hline
H$^{13}$CN    & (2-1)       & 42.7$\pm$6.4  & 72.4$\pm$10.9  & 95.7$\pm$14.4  & 35.0$\pm$5.2  \\
DCN      &(2-1)        & 1.9$\pm$0.3   & 2.3$\pm$0.3    & 3.4$\pm$0.5    & 1.0$\pm$0.1   \\
N$_2$H$^+$     & (2-1)        & 82.7$\pm$12.4 & 191.0$\pm$28.6 & 173.0$\pm$25.9 & 99.1$\pm$14.9 \\
H$^{13}$CO$^+$   & (2-1)        & 47.8$\pm$7.2  & 67.6$\pm$10.1  & 88.6$\pm$13.3  & 16.6$\pm$2.5    \\ \hline
\end{tabular}

\end{table*}

\begin{table*}
\centering
\caption{The deuterium fractionations for each GMC region shown for DCN as well as the 1 sigma upper limits obtained for DCO$^+$ and N$_2$D$^+$ as a result of the LTE fitting conducted using \texttt{MADCUBA}, the T$_{\text{ex}}$ for each region are also provided.}
\label{tab:Ratios}
\begin{tabular}{ccccccc}
\hline
Region & DCN/HCN  & DCO$^+$/HCO$^+$  & N$_2$D$^+$/N$_2$H$^+$ & $T_{\text{ex}}$ & FWHM & \twelve C/\thirteen C \\ & ($\times 10^{-4}$) & ($\times 10^{-4}$) & ($\times 10^{-3}$) & (K) & (km s$^{-1}$ & \\ \hline
GMC-3  & 6.43 $\pm$ 0.74           & ($<1.51$)                         & ($<4.47$)  & 50 $\pm$ 5 & 74 & 48 $\pm$ 4                   \\
GMC-4  & 5.58 $\pm$ 1.06          & ($<0.8$)                          & ($<2.51$)  & 60 $\pm$ 7 & 74 &  57 $\pm$ 10                   \\
GMC-6  & 10.3 $\pm$ 0.1           & ($<1.08$)                         & ($<3.02$)     & 98 $\pm$ 4 & 74  &  41 $\pm$ 2               \\
GMC-7  & 6.75 $\pm$ 1.23          & ($<5.49$)                         & ($<4.17$)         & 61 $\pm$ 11 & 79 & 44 $\pm$ 5     \\ \hline
\end{tabular}
\end{table*}

\section{Discussion and Conclusions}
\label{sec:Disc_Conc}
Observing deuterated species in  extragalactic environments is key to understanding deuteration processes within the ISM of galaxies. The degrees of fractionation observed in DCN/HCN in NGC 253 are consistent with similar galactic CMZ studies covering regions of similar kinetic temperatures \citep[e.g.][]{2022_Colzi}.

In order to properly interpret our results we shall compare our observed values and upper limits of D/H to those observed in other sources. In Figure \ref{fig:Swarm}, we show how the D/H ratio from column density ratios and the line intensity ratios of DCN (2-1)/H\thirteen CN (2-1) compare to the recent galactic D/H ratio obtained from DCN by \cite{2022_Colzi}. 
 The intensity and column density ratios as shown in Figure \ref{fig:Swarm} generally agree in each of the GMCs.
 \cite{2022_Colzi} observed two components within G+0.693-0.027, a molecular cloud located within the Sgr B2 complex of the Milky Way's central molecular zone. Using Non-LTE Radiative Transfer modelling they were able to identify physical conditions of a prestellar (PS) component (T$_K \sim 30$K) and a broad warm component (T$_K \sim 100$K). These components were observed with varying deuterium fractionations as observed with isotopologues of dense gas tracers, such as HCN, HCO\plus and HNC.
When comparing to the results of \cite{2022_Colzi} it can be seen that GMC-6 is quite comparable to the broad, warmer component of G+0.693-0.027 . The D/H ratio of $(10.3 \pm 0.1) \times 10^{-4}$ we observe towards GMC-6 is comparable to the $(8.6-13)\times 10^{-4}$ observed in G+0.693-0.027's broad component by \cite{2022_Colzi} also using DCN. The broad component of G+0.693-0.027 and the gas component traced by DCN and H\thirteen CN, within GMC-6, in this project both also seem to have comparable kinetic temperatures ($\sim 100$ K). GMCs -4,-5 and -7 on the other hand each have DCN/HCN fractions (D/H = $5.58 - 6.75\times 10^{-4}$ which lie between that observed by the cooler prestellar component, D/H $\sim4 \times 10^{-4}$) and the broad component of G+0.693-0.027, D/H $=(8.6-13)\times 10^{-4}$. This is consistent with the observed temperatures of these regions ($\sim60$ K, shown in Table \ref{tab:Ratios} which are predicted to both being warmer than the cooler prestellar region of G+0.693-0.027 but less warm than the broad warm component. This all implies a similarity between the CMZ of NGC~253 and our own galaxy. The upper limit D/H ratio as derived from a tentative detection of DCO\plus is  $<8 \times 10^{-5}$, within GMC-4, which is relatively similar to the cosmic value of $\sim2 \times 10^{-5}$ \citep{2018Zavarygin}.  It is also important to note that \cite{2022_Colzi} detected similarly low D/H ratios within the broad component of G+0.693-0.027 through N$_2$D\plus and DCO\plus and their hydrogenated counterparts. 

When comparing our deuterium fractions to those observed within the LMC using DCN by \cite{Chin1996} we see that they appear to be low relative to their $\sim2 \times 10^{-3}$. It should be noted that the LMC is a known low metallicity environment and thus this may contribute to the relatively high deuterium fraction observed, since the LMC gas is less contaminated by stellar processing, negatively affecting the deuterium abundance. 

 Finally, we compare our D/H values to those observed in a shocked region. The shocked molecular outflow region L1157-B1 was observed by \cite{2017_Busquet} to have D/H values of the of an order higher than observed within this study at D/H$\sim 4 \times 10^{-3}$. The predicted temperatures of this region of ~80 K are also relatively consistent with those predicted for the regions of this study. As can be seen in Figure \ref{fig:Swarm} the observed D/H fraction within this region is significantly higher, implying an enhancement on the deuterium fractionation by shocks.
    
It is also important to compare our observed D/H ratios to those predicted by appropriate chemical modelling. Figure \ref{fig:Model_DCN} shows how the range of D/H ratios across the GMC regions compares to previous investigations of this ratio in DCN/HCN using chemical models from \cite{2007_Roueff} and \cite{Bayet2010}. In theory the most appropriate model for the conditions of GMCs within the CMZ of NGC 253 would be that of the cosmic-ray enhanced starburst environments (denoted as `SB+') from \cite{Bayet2010}, specifically the $10^6$ cm$^{-3}$ density case \citep[similar to the density derived by][]{2019_Mangum_Heart}. The `SB+' is in full agreement with the enhanced cosmic ray ionization rate towards the central molecular zone of NGC 253 probed by various molecular species \citep{2021Holdship,2021Harada,2022Holdship,2022Behrens}. As can be seen in Figure \ref{fig:Model_DCN} however, this model significantly over predicts the DCN/HCN ratio relative to what we have observed. The D/H ratio appears to increase in GMC-6 relative to the other 3 regions and this also corresponds to a significantly higher temperature in GMC-6 relative to the other GMCs; a plot showing this trend can be seen in Figure \ref{fig:DCN_Temp}. The `SB+' model in \cite{Bayet2010} was run with a temperature of 300 K  which is significantly higher than the temperature range we  probe with these observations  ($\sim$50-100K). DCN is favoured to form from HD present in warmer gas and so this may explain the overprediction in the \cite{Bayet2010} models. The models from \cite{2007_Roueff} were run using more comparable temperatures (50-70 K) and as can be seen in Figure \ref{fig:Model_DCN} the warm core model at a density of $10^6$cm$^{-3}$ and temperature of 70 K fits very well with our observations, as they also do to the broad component of G+0.693-0.027 studied by \cite{2022_Colzi}. This may also suggest that the cosmic ray ionisation rate is not important for deuterium fractionation processes.

Approaching the warmer temperatures of the gas traced by these molecules (> 60 K) the most efficient formation mechanism for DCN in the ISM is that of CH$_2$D\plus reacting with atomic N \citep{2007_Roueff,2013_Roueff}. The CH$_2$D\plus is formed by the exothermic reaction
\begin{equation}
    \text{CH}_3^+ + \text{HD} \rightarrow \text{CH}_2\text{D}^+ + \text{H}_2,
\end{equation}
which becomes the primary processing reaction of HD (the most abundant deuterated species) at >50 K. The CH$_2$D\plus then reacts with the atomic N, in the following reaction
\begin{equation}
    \text{CH}_2\text{D}^+ + \text{N} \rightarrow \text{DCN}^+ + \text{H}_2.
\end{equation}
The DCN$^+$ cation then undergoes hydrogenation and dissociative recombination reactions to form DCN. \cite{2007_Roueff} performed chemical models at 50-70 K and observed D/H ratios in DCN and HCN of $>5 \times 10^{-4}$, which are consistent with the observed results of this paper.

In conclusion the results of this paper are:
\begin{itemize}
\item Observing DCN within the CMZ of NGC253 is advantageous  thanks to the relatively close proximity of this galaxy. Our observation provides us with the opportunity to study deuteration in star-forming regions at cloud-scale resolution despite the limited spatial resolution inherent to extragalactic studies. As described in Section \ref{sec:Int}, NGC 253 contains multiple well observed GMCs as well as in Super Star Clusters (SSCs).

\item We were able to constrain D/H ratio estimates within 4 regions of the CMZ of NGC~253 via  isotopologues of HCN. We obtained a range between 5.58 $\pm$ 1.06 $\times 10^{-4}$ and 10.3 $\pm$ 0.1 $\times 10^{-4}$ across the 4 GMCs. The warmest observed region, GMC-6, comparing favourably to the warm, broad component of G+0.693-0.027 observed within the CMZ of our Galaxy. The remaining 3 GMCs, each with cooler predicted temperatures aligned more closely to the cooler prestellar component observed within G+0.693-0.027 as observed by \cite{2022_Colzi}. 
\item Our results seem to be consistent with the idea that warmer temperature gas components lead to an increase in the abundance of DCN.

\item In order to be more certain of the exact conditions of the gas being traced by DCN, other species such as N$_2$D$^+$ and DCO\plus need to be detected, as thanks to their preference of lower temperature formation mechanisms the correlation between deuterium fractionation and temperature could be better constrained in these extragalactic studies.
\end{itemize}

To conclude, in this paper we have shown the first detection of a deuterated molecule in an extragalactic starburst environment, beside the LMC. This study provides another step in the progression of the study of deuterated species in the ISM by providing a comparison to the numerous observations of deuterated molecules within our own Galaxy.


\section*{Acknowledgements}

This paper makes use of the following ALMA
data: ADS/JAO.ALMA$\#$2017.1.00028.S ALMA is a partnership
of ESO (representing its member states), NSF (USA) and NINS (Japan), together with NRC (Canada), NSTC and ASIAA (Taiwan), and KASI (Republic of Korea), in co-operation with the Republic of Chile. The Joint ALMA Observatory is operated by ESO, AUI/NRAO and NAOJ. J.B. and S.V. have received funding from the European Research Council (ERC) under the European Union’s Horizon 2020 research and innovation programme MOPPEX 833460.
V.M.R. and LC acknowledge support from the grants No. PID2019-105552RB-C41 and PID2022-136814NB-I00 by the Spanish Ministry of Science, Innovation and Universities/State Agency of Research MICIU/AEI/10.13039/501100011033 and by "ERDF A way of making Europe". VMR also acknowledges support from the grant RYC2020-029387-I funded by MICIU/AEI/10.13039/501100011033 and by "ESF, Investing in your future", from the Consejo Superior de Investigaciones Cient{\'i}ficas (CSIC) and the Centro de Astrobiolog{\'i}a (CAB) through the project 20225AT015 (Proyectos intramurales especiales del CSIC); and from the grant CNS2023-144464 funded by MICIU/AEI/10.13039/501100011033 and by “European Union NextGenerationEU/PRTR”. The authors would like to thank the anonymous referee for the constructive comments that greatly increased the quality of this paper from its original version.

%
%

\bibliographystyle{aa}
\bibliography{Bib}

\begin{appendix}

\FloatBarrier

\section{Additional Spectral Information}
\label{app:Spec_Info}
\begin{table}
\centering
\caption{Spectroscopy of the HC$_3$N and its isotopologues studied in this work from CDMS \citep{CDMS_2001,CDMS_2005,CDMS_2016}.}
\begin{tabular}{c c c c c}
\hline
Molecule & Frequency & Transition  & logA$_{\rm ul}$  & E$_{\rm up}$  \\
 & (GHz) &   &  (s$^{-1}$) & (K)  \\
\hline
HC$_3$N & 154.657284 & 17$-$16 & -3.53575  & 66.8   \\
HC$_3$N & 172.8493 & 19$-$18 & -3.3896  & 83.0   \\
\hline
HCC\thirteen CN& 144.957454 & 16$-$15 & -3.62088  &  59.1  \\
HCC\thirteen CN& 154.016078 & 17$-$16 & -3.54113  &  66.5  \\
\hline
HC\thirteen CCN& 144.943467 & 16$-$15 &  -3.62109 & 59.1   \\
HC\thirteen CCN & 154.001217 & 17$-$16  & -3.54134  & 66.5   \\
\hline 
\end{tabular}
\label{tab:transitions2}
\end{table}
\newpage
\FloatBarrier
\section{Additional Moment 0 Maps}
\label{app:Mom0maps}
\begin{figure}
\centering
  \begin{tabular}[b]{@{}p{1.0\linewidth}@{}}
    \centering\includegraphics[width=1.0\linewidth]{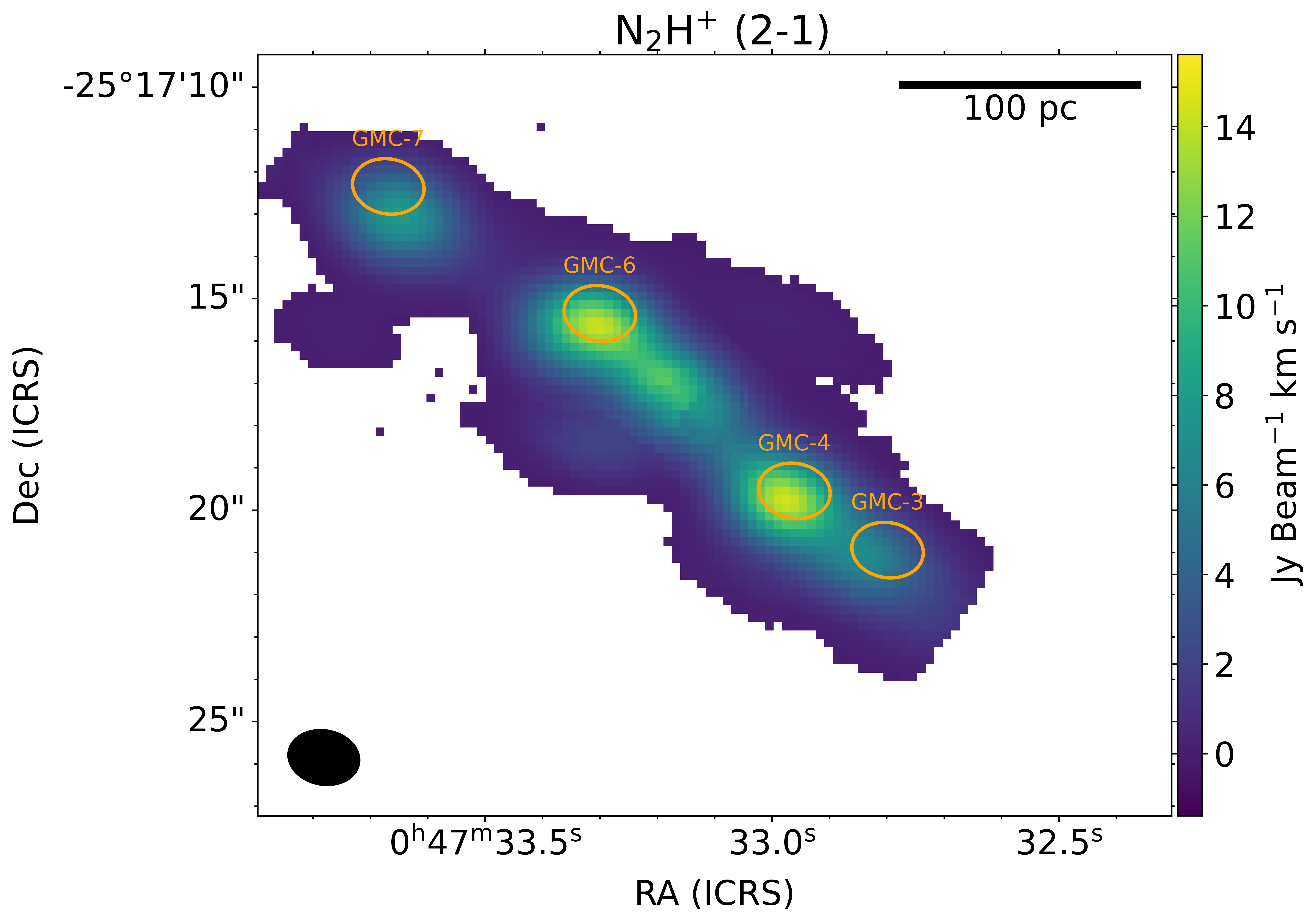} 

    \centering\small (a)
      \end{tabular}%
  \quad
  \begin{tabular}[b]{@{}p{1.0\linewidth}@{}}
    \centering\includegraphics[width=1.0\linewidth]{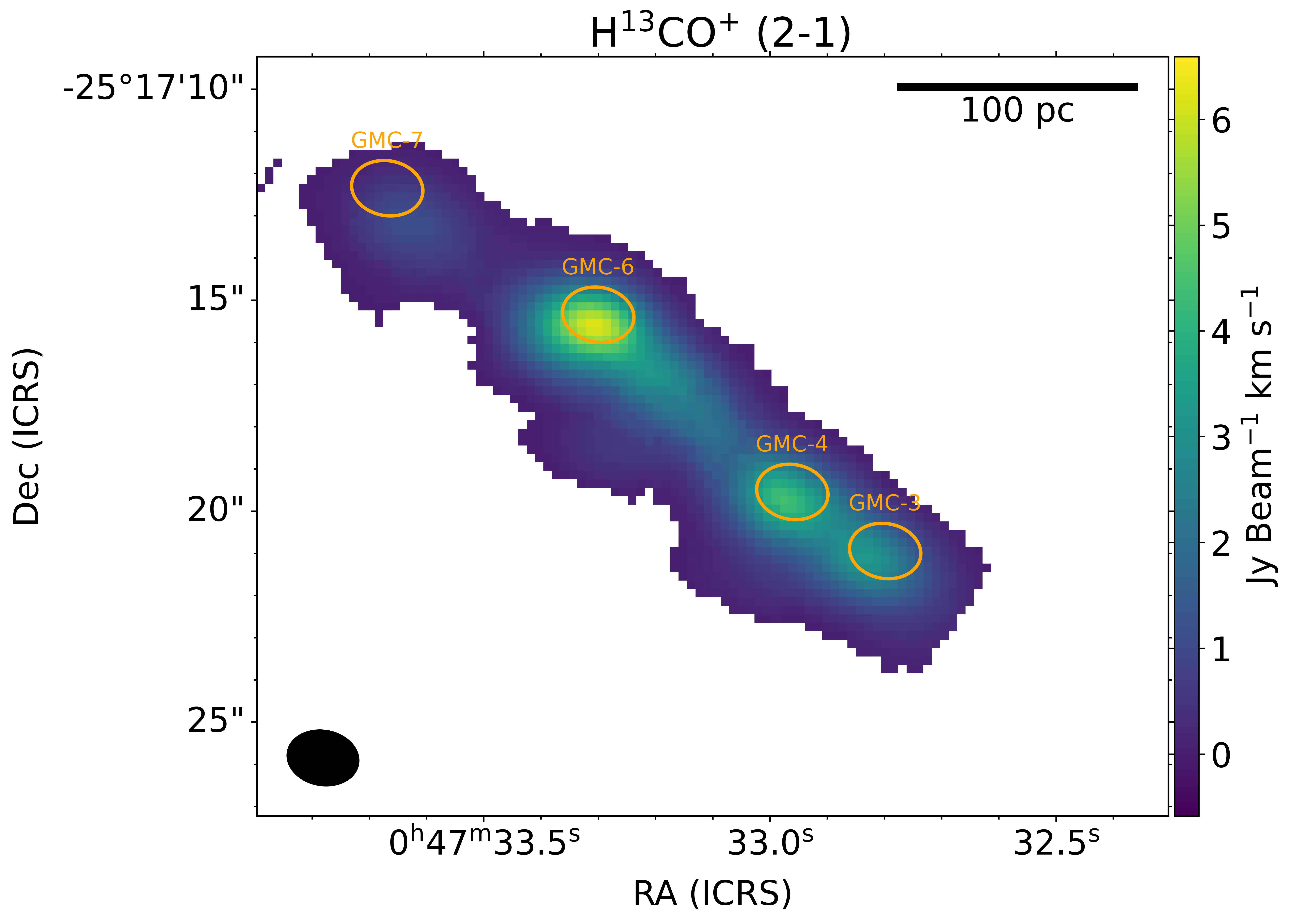} 
    \centering\small (b) 
  \end{tabular}%
\caption{Velocity-integrated line intensity moment 0 maps, given in [Jy\,km\,s$^{-1}$ /beam], of the N$_2$H\plus (2-1) and H\thirteen CO\plus (2-1) lines observed in the CMZ of NGC 253.
  }
  \label{fig:mom0_others}
\end{figure}

\begin{figure}
\centering
  \begin{tabular}[b]{@{}p{1.0\linewidth}@{}}
    \centering\includegraphics[width=1.0\linewidth]{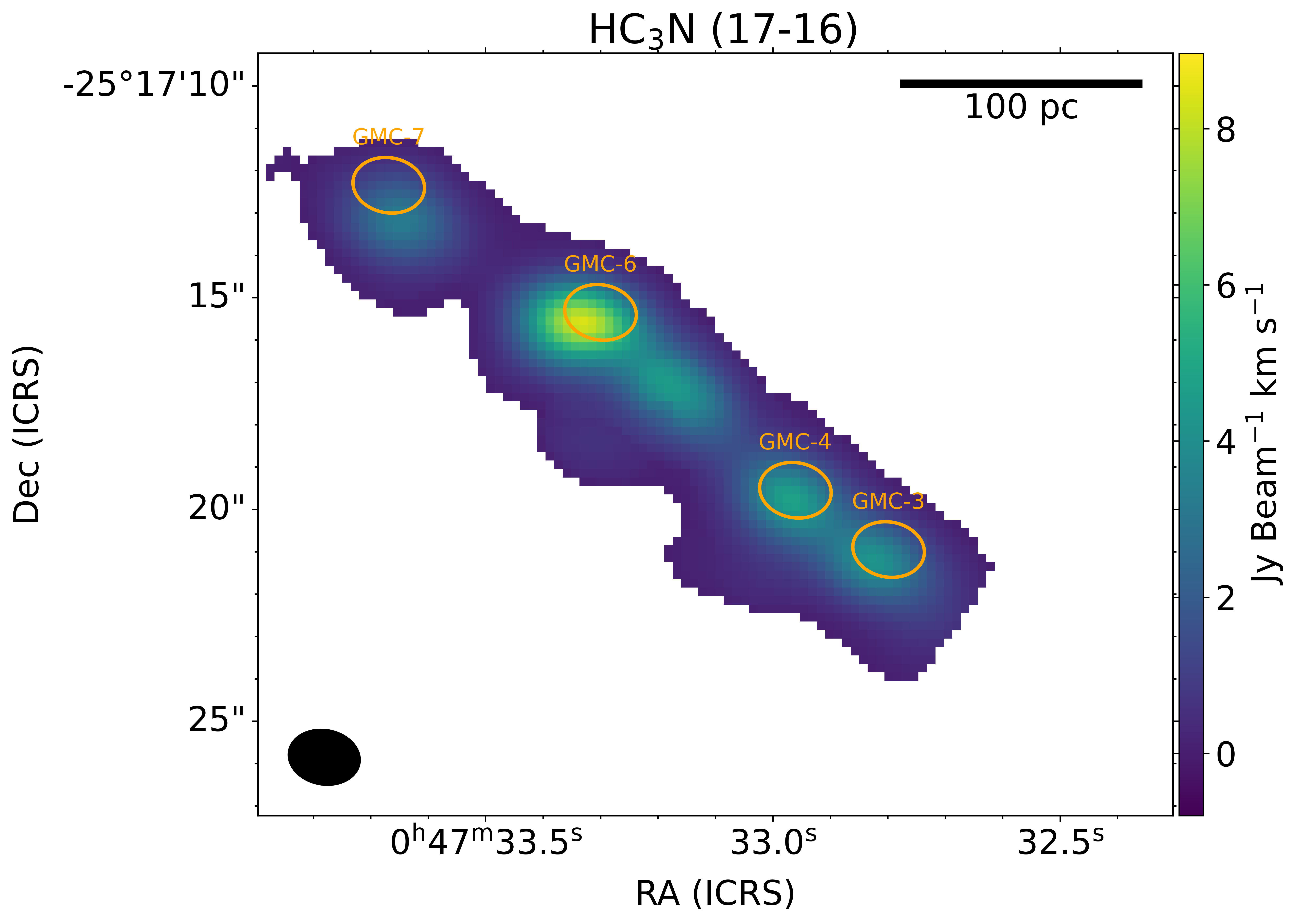} 

    \centering\small (a)
      \end{tabular}%
  \quad
  \begin{tabular}[b]{@{}p{1.0\linewidth}@{}}
    \centering\includegraphics[width=1.0\linewidth]{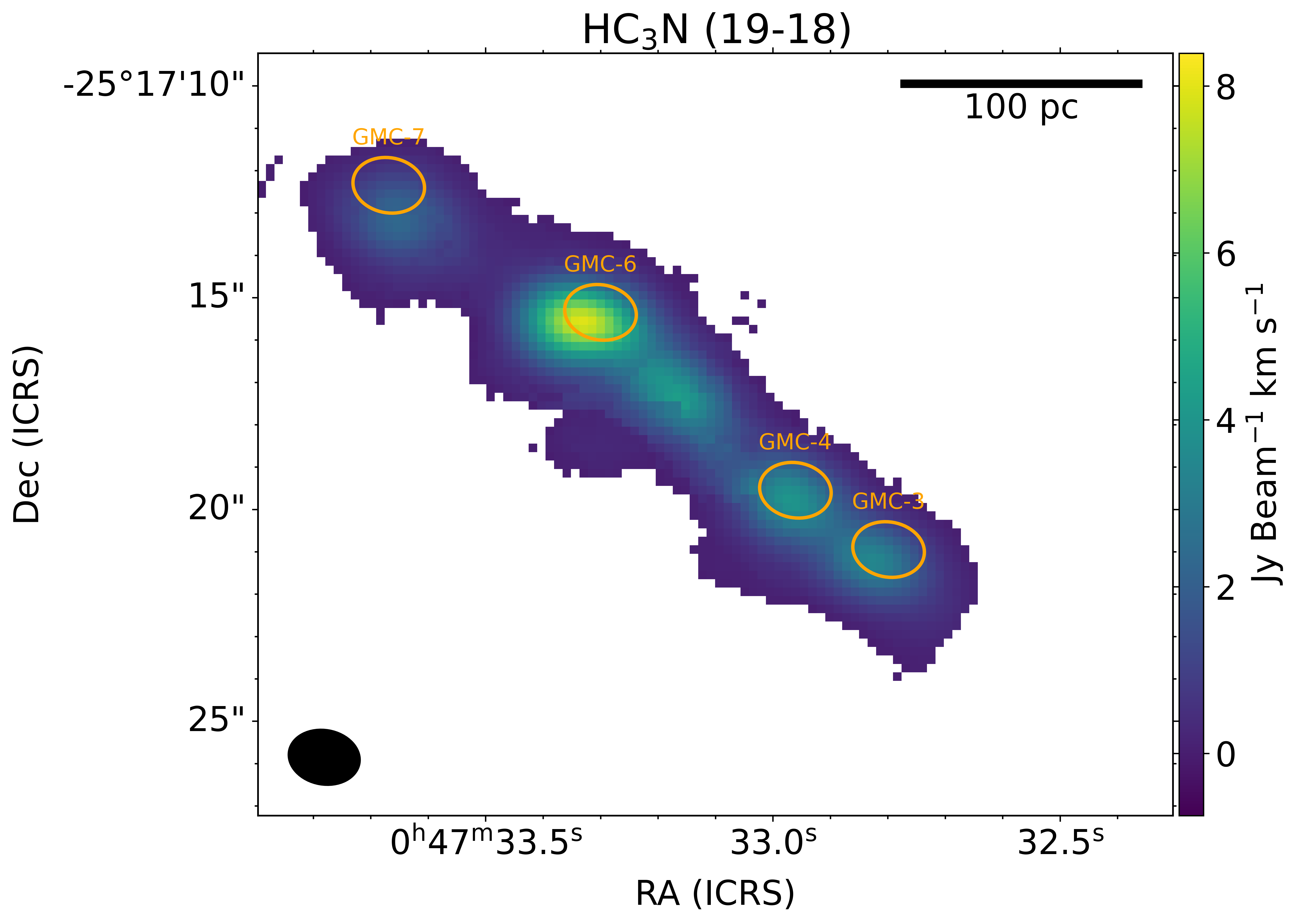} 
    \centering\small (b) 
  \end{tabular}%
\caption{Velocity-integrated line intensity moment 0 maps, given in [Jy\,km\,s$^{-1}$ /beam], of the HC$_3$N (17-16) and HC$_3$N (19-18) lines observed in the CMZ of NGC 253. 
  }
  \label{fig:mom0_HC3N}
\end{figure}

\newpage
\FloatBarrier

\section{Spectra}
\label{app:Spectra}

\begin{figure*}
    \centering
    \includegraphics[width=1.0\textwidth]{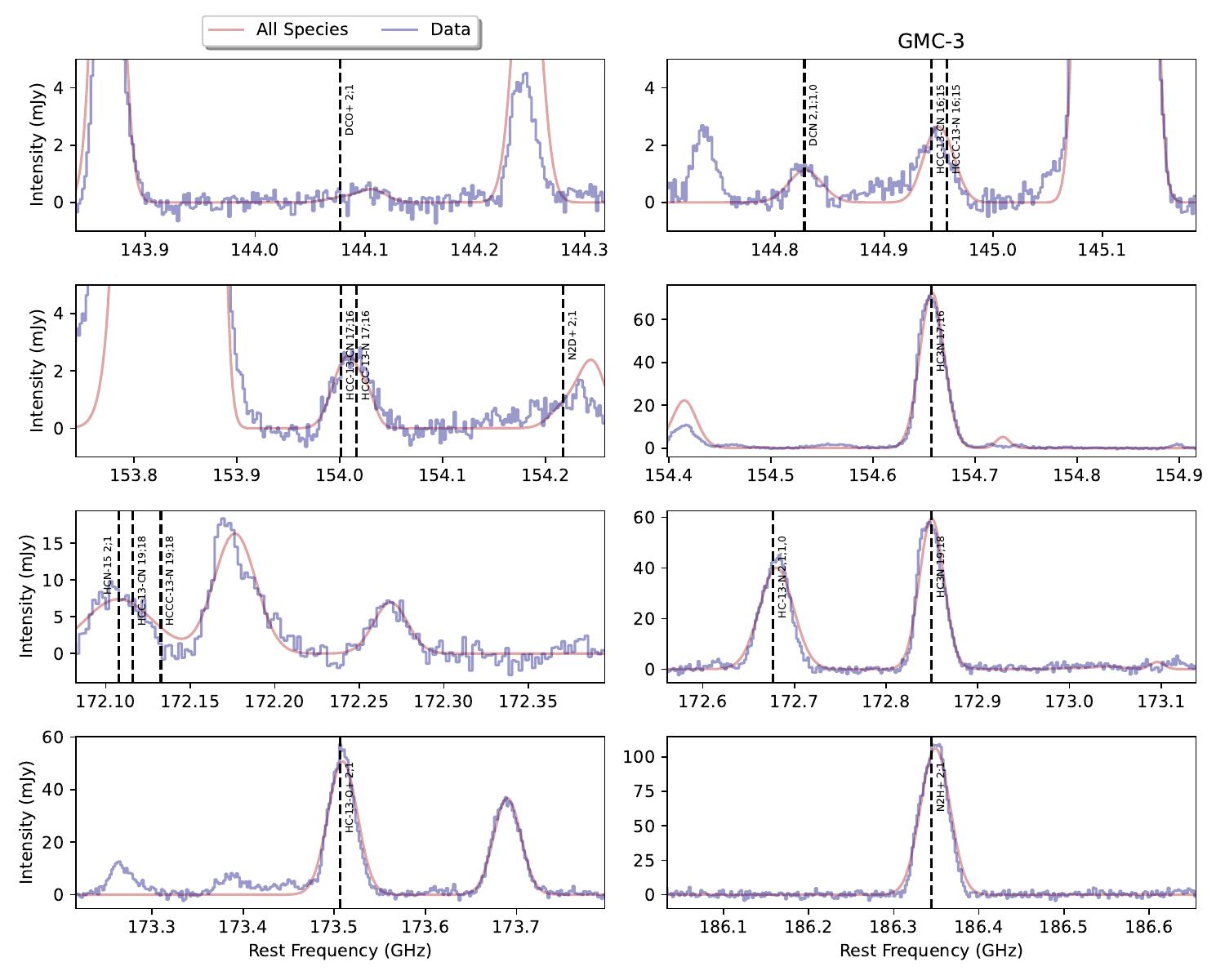}
    \caption{The spectra of selected lines observed in the region GMC-3 (in blue). The best fitting LTE model from \texttt{MADCUBA} is shown overplotted in red.}
    \label{fig:GMC3_Spectra}
\end{figure*}

\begin{figure*}
    \centering
    \includegraphics[width=1.0\textwidth]{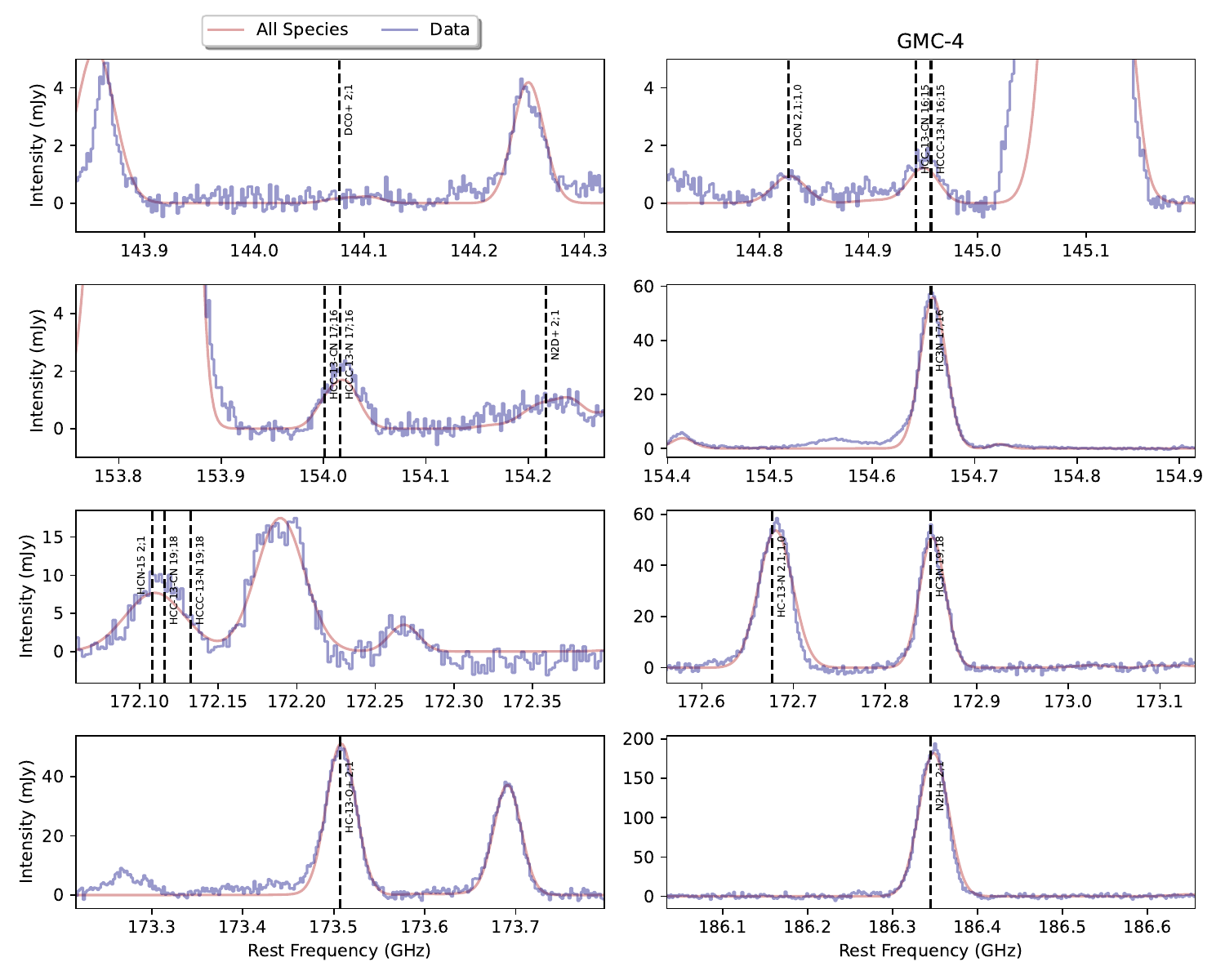}
    \caption{The spectra of selected lines observed in the region GMC-4 (in blue). The best fitting LTE model from \texttt{MADCUBA} is shown overplotted in red.}
    \label{fig:GMC4_Spectra}
\end{figure*}

\begin{figure*}
    \centering
    \includegraphics[width=1.0\textwidth]{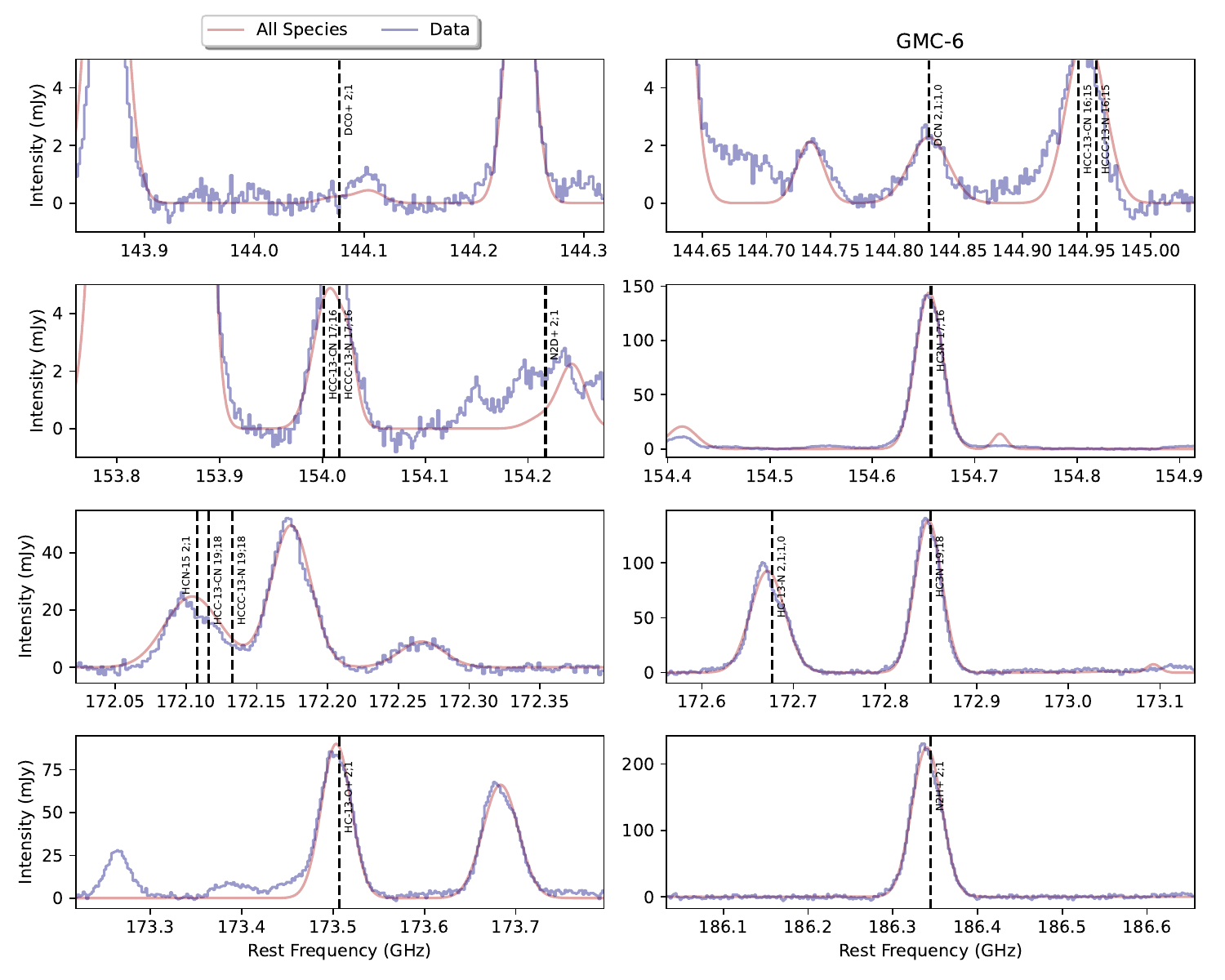}
    \caption{The spectra of selected lines observed in the region GMC-6 (in blue). The best fitting LTE model from \texttt{MADCUBA} is shown overplotted in red.}
    \label{fig:GMC6_Spectra}
\end{figure*}

\begin{figure*}
    \centering
    \includegraphics[width=1.0\textwidth]{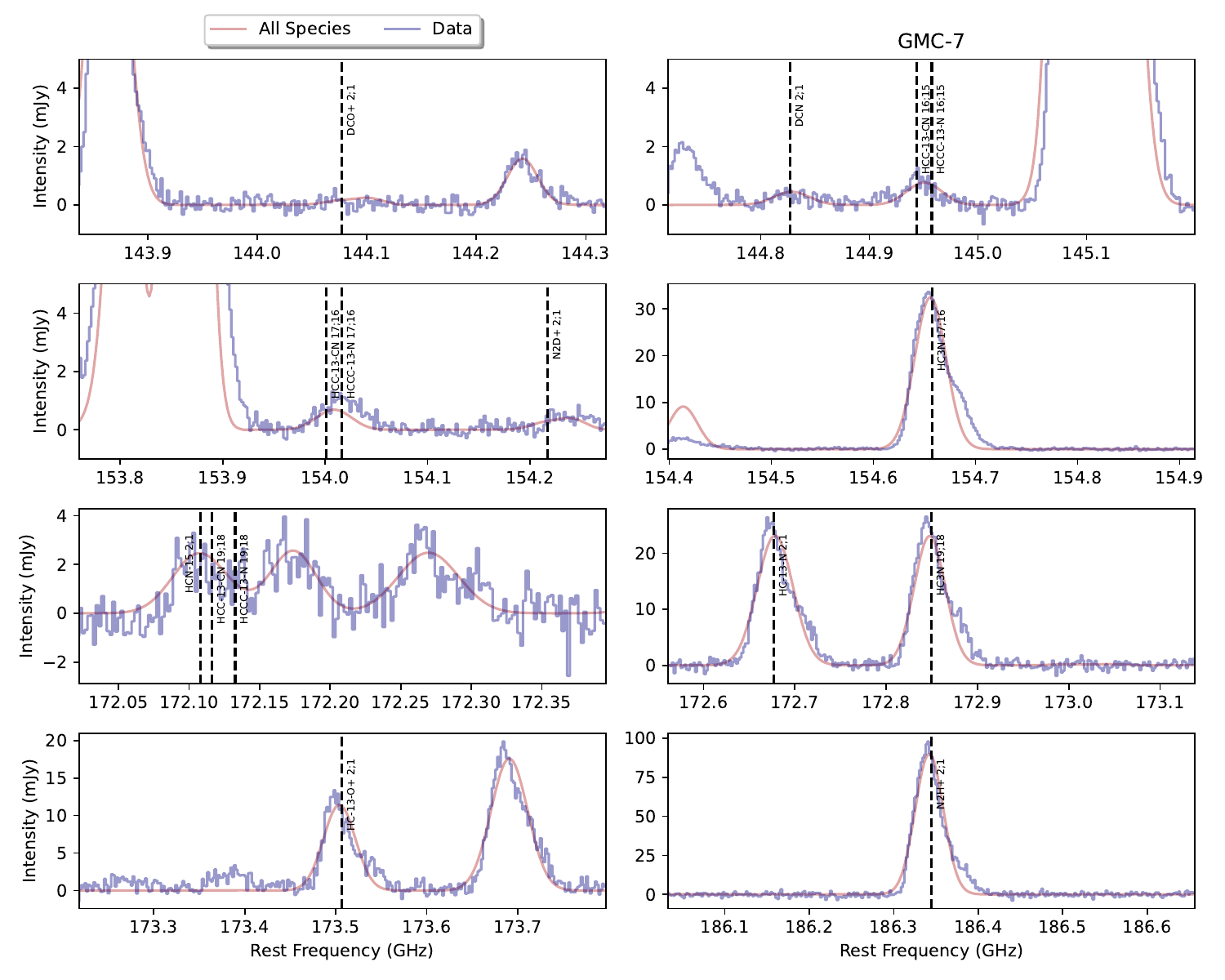}
    \caption{The spectra of selected lines observed in the region GMC-7 (in blue). The best fitting LTE model from \texttt{MADCUBA} is shown overplotted in red.}
    \label{fig:GMC7_Spectra}
\end{figure*}
\newpage
\FloatBarrier

\end{appendix}
\end{document}